\documentclass[12pt,preprint]{aastex63}
\usepackage{natbib}
\usepackage{url}

\shorttitle{Pre-White Dwarf Companion of Regulus} 
\shortauthors{Gies et al.} 
\input{epsf} 

\begin{document} 
 
\title{Spectroscopic Detection of the Pre-White Dwarf Companion of Regulus} 

\correspondingauthor{Douglas Gies}
\email{gies@chara.gsu.edu, lester@astro.gsu.edu, lwang@chara.gsu.edu, andcoup@astro.gsu.edu, shepard@astro.gsu.edu,}
\email{Coralie.Neiner@obspm.fr, Gregg.Wade@rmc.ca, dunham@starpower.net, Business@occultations.org}

\author[0000-0001-8537-3583]{Douglas R. Gies}
\affiliation{Center for High Angular Resolution Astronomy and
 Department of Physics and Astronomy,\\
 Georgia State University, P. O. Box 5060, Atlanta, GA 30302-5060, USA}

\author[0000-0002-9903-9911]{Kathryn V. Lester}
\affiliation{Center for High Angular Resolution Astronomy and
 Department of Physics and Astronomy,\\
 Georgia State University, P. O. Box 5060, Atlanta, GA 30302-5060, USA}

\author[0000-0003-4511-6800]{Luqian Wang}
\affil{Center for High Angular Resolution Astronomy and
 Department of Physics and Astronomy,\\
 Georgia State University, P. O. Box 5060, Atlanta, GA 30302-5060, USA}

\author[0000-0001-9834-5792]{Andrew Couperus}
\affil{Center for High Angular Resolution Astronomy and
 Department of Physics and Astronomy,\\
 Georgia State University, P. O. Box 5060, Atlanta, GA 30302-5060, USA}
 
\author[0000-0003-2075-5227]{Katherine Shepard}
\affiliation{Center for High Angular Resolution Astronomy and  
 Department of Physics and Astronomy,\\ 
 Georgia State University, P. O. Box 5060, Atlanta, GA 30302-5060, USA}


\author[0000-0003-1978-9809]{Coralie Neiner}
\affiliation{LESIA, Observatoire de Paris, PSL University, CNRS, 
Sorbonne Universit\'{e}, Univ. Paris Diderot, \\
Sorbonne Paris Cit\'{e}, 5 place Jules Janssen, F-92195 Meudon, France}

\author[0000-0002-1854-0131]{Gregg A. Wade}
\affiliation{
Department of Physics and Space Science, Royal Military College of Canada, 
P.O.\ Box 17000 Station Forces, Kingston, ON K7K 0C6, Canada}

\author[0000-0001-7527-4207]{David W. Dunham}
\affiliation{IOTA, 7913 Kara Court, Greenbelt, MD 20770, USA}

\author{Joan B. Dunham}
\affiliation{IOTA, 7913 Kara Court, Greenbelt, MD 20770, USA}

 
 
\begin{abstract} 
Mass transfer in an interacting binary will often strip the mass donor
of its entire envelope and spin up the mass gainer to near critical rotation. 
The nearby B-type star Regulus represents a binary in the post-mass 
transfer stage: it is a rapid rotator with a very faint companion in 
a 40 d orbit. Here we present the results of a search for 
the spectral features of the stripped-down star in an extensive 
set of high S/N and high resolution spectra obtained with 
the CFHT/ESPaDOnS and TBL/NARVAL spectrographs.  We first determine 
revised orbital elements in order to set accurate estimates of the 
orbital Doppler shifts at the times of observation.  We then calculate 
cross-correlation functions of the observed and model spectra, and 
we search for evidence of the companion signal in the residuals after
removal of the strong primary component.  We detect a weak peak in 
the co-added residuals that has the properties expected for a faint 
pre-white dwarf.  We use the dependence of the peak height and width on 
assumed secondary velocity semiamplitude to derive the semiamplitude,
which yields masses of $M_1/M_\odot = 3.7 \pm 1.4$ and $M_2/M_\odot = 0.31 \pm 0.10$
(assuming orbital inclination equals the spin inclination of Regulus).
We estimate the pre-white dwarf temperature $T_{\rm eff} = (20 \pm 4)$~kK
through tests with differing temperature model spectra, and we find the 
radius $R_2/R_\odot = 0.061 \pm 0.011$ from the component temperatures 
and the flux ratio associated with the amplitude of the signal in the 
cross-correlation residuals. 
\end{abstract} 
 
 

\section{Introduction}                              
 
Stars born with close by neighbors are destined to interact in ways 
that will dramatically change their future paths \citep{DeMarco2017}. 
The more massive component will be the first to grow to dimensions 
comparable to the separation, and subsequent Roche lobe overflow will lead to 
the transfer of mass and angular momentum to the mass gainer star. 
This process often leads to orbital shrinkage until the mass ratio 
is reversed, at which point continuing mass transfer causes the orbit 
to re-expand (if the stars avoid a common envelope stage or a merger). 
The result will be a system composed of a rapidly rotating, 
main sequence star and a stripped-down, stellar core \citep{Willems2004}. 
Detailed evolutionary sequences with the MESA code show that the 
remnants of massive systems may produce stripped cores massive enough
to create H-free supernovae \citep{Gotberg2018} or very low mass, 
white dwarf progenitors for intermediate mass binaries \citep{Chen2017}. 
This post-mass transfer stage can be relatively long-lasting, so 
many such systems should exist, but they are hard to detect because the 
faint stripped core is lost in the glare of the main sequence star.  

Detailed observational work is now beginning to uncover this hidden 
population of post-mass transfer binaries.  \citet{Pols1991} suggested 
that the rapidly rotating, B-emission line stars (i.e., classical Be stars) 
were spun up through mass transfer, and investigations of their ultraviolet 
spectra have revealed the spectral signatures of hot, stripped companions in a 
growing number of cases \citep{Wang2018}.  High precision photometric 
surveys from ground- and space-based observations have detected the 
eclipses of hot cores by intermediate mass stars that form the EL~CVn 
class of binaries \citep{Maxted2014,Rappaport2015,vanRoestel2018,Wang2020}.
The main sequence components are boosted in mass and luminosity 
through mass transfer, and they may appear as ``blue straggler'' stars 
lying above the turn-off point in the color-magnitude diagrams of clusters. 
\citet{Gosnell2019} found that the hot cores of the mass donors 
can often be detected through ultraviolet spectroscopy of cluster blue 
stragglers because the white dwarf flux dominates over that of the cooler 
companion at short wavelength. 

Stars with very rapid rotation are good candidates for components spun up
by prior mass transfer in a binary.  One of the brightest of the rapid rotators 
is the star Regulus ($\alpha$ Leo; HD~87901; HR~3982; HIP~49699), which 
is a nearby ($d=24.3\pm0.2$ pc; \citealt{vanLeeuwen2007}), intermediate mass 
star with a spectral classification B8~IVn \citep{Gray2003}.  Images of 
Regulus from interferometry with the CHARA Array show that it has a 
rotationally distorted figure that we view towards its equator
\citep{McAlister2005,Che2011}.  \citet{Gies2008} presented a large number 
of radial velocity measurements from archival spectra of Regulus, and 
they found that the star is the brighter member of a binary with a 40~d 
orbital period.  The low value of the orbital mass function and the lack 
of any evidence of spectral features from the companion suggested that the 
companion is a low mass and faint pre-white dwarf, the remnant core of the 
former mass donor in the binary.  \citet{Rappaport2009} showed that the 
orbital period and probable mass of the suspected white dwarf match the 
expected relation for stars that lose their envelopes during the red giant
phase, and they developed a detailed model of the past and future evolutionary
paths of the binary system. 

The hunt to find the flux of the white dwarf companion led to a possible detection
on 2016 October 13 in Papua New Guinea when David and Joan Dunham observed 
an occultation of Regulus by the asteroid Adorea 
\citep{Dunham2017}\footnote{http://www.occultations.org/meetings/NA/2017Meeting/Dunham\_RegulusAdorea.ppt}.  
They recorded the event with a 10-inch ``suitcase telescope'' and video camera, 
and their results detected the presence of a faint ($V\approx 12$ mag) star
at the precise location of Regulus during the 3~s occultation of Regulus itself. 
The chance is remote that a field star this bright would be found exactly at 
the position of Regulus, so they suggested that the faint star is the binary 
companion that escaped occultation by Adorea. 

A spectroscopic detection of the companion of Regulus would certainly 
advance our knowledge of the system parameters and evolutionary state. 
Identifying the spectral features of the faint companion would yield new 
information on the orbital Doppler shifts and masses as well as the pre-white 
dwarf temperature and flux contribution.  Here we present 
an analysis of additional archival spectroscopy of Regulus with the goal
of detecting the spectral signature of the pre-white dwarf.  Section 2 describes 
the additional spectroscopy data that we measured to derive radial velocities
and improved orbital elements.  We then focus in Section 3 on an homogeneous
set of high resolution and high S/N echelle spectra that is ideal for 
extracting the faint signal of the pre-white dwarf spectrum.  We conclude 
in Section 4 with a discussion of the observational results and their
implications for our understanding of the evolutionary state of both components.
 
 
\section{Radial Velocities and Revised Orbital Elements} 

The first step in a search for the companion is to verify the orbital elements
in order to estimate the companion's velocity at the time of the observations. 
In our earlier paper on Regulus \citep{Gies2008}, we presented radial velocity 
measurements based upon 15 sets of spectroscopic data as listed in Table~1 
of that paper.  Here we present new measurements from five additional sets 
of spectra that supplement the original sets.  Table 1 lists the 
observational dates and characteristics of the spectrograph for each of these
with an incremental number assigned to each source that continues from the 
original work. The data sources include the following: \\
\#16.\ Echelle spectra obtained with the Okayama Astrophysical Observatory 
1.88 m telescope (Japan) and fiber-fed HIDES spectrograph \citep{Kambe2013}.  
The radial velocities were measured for the H$\beta$ line with the cross-correlation 
technique using the {\tt rv.fxcor} task in IRAF \citep{Alpaslan2009}.
This set is based upon 139 spectra, and the results obtained close in time 
were combined to form 17 nightly averages and their standard deviations. \\
\#17.\ Spectra made with the Bochum \'{E}chelle Spectroscopic Observer (BESO) 
fiber-fed, high-resolution spectrograph on the 1.5-m Hexapod Telescope at the 
Cerro Armazones Observatory (Chile) \citep{Fuhrmann2011}.  These measurements were 
summarized by \citet{Chini2012} and include five spectra for three nightly averages. \\
\#18.\ Spectra from the California Planet Survey made with the HIRES echelle 
spectrograph on the Keck~I telescope.  The measurements were presented by 
\citet{Becker2015} and are based upon 21 spectra yielding four nightly averages. \\
\#19.\ Ultraviolet spectra obtained as part of the {\it HST} STIS Advanced Spectral 
Library (ASTRAL) Project \citep{Ayres2014}.  This set includes six E140H and eight 
E230H echelle spectra that were obtained in three contiguous time groups spread 
over four days.  These were re-binned to a resolving power of $R = 20000$ on a 
$\log \lambda$ grid, and then each spectrum was cross-correlated with a model spectrum for 
Regulus from the UVBLUE grid \citep{Rodriguez2005}.  Radial velocities from the 
cross-correlation maxima positions were averaged and standard deviations determined 
for each of the three groups obtained close in time. \\
\#20.\ This is a remarkable set of spectra hosted at the {\it PolarBase} 
web site\footnote{http://polarbase.irap.omp.eu/} 
\citep{Petit2014} that is an archive of data from the ESPaDOnS and NARVAL high-resolution 
spectropolarimeters on the Canada-France-Hawaii Telescope (CFHT) 
and the T\'{e}lescope Bernard Lyot (TBL, Pic du Midi Observatory), respectively.  
Most of these were obtained as part of 
the {\it MiMeS} (Magnetism in Massive Stars) survey \citep{Wade2016} and 
the {\it BritePol} survey \citep{Neiner2017}.  We collected 786 spectra that were 
re-binned onto a standard $\log \lambda$ wavelength grid with a lower resolving power 
of $R=20000$ in order to increase the spectral S/N, and then 
nightly average spectra were formed for 25 dates.  These spectra were cross-correlated 
with a model spectrum for Regulus from the BLUERED grid \citep{Bertone2008} over the range from 
3941 to 4936 \AA , where the signal-to-noise ratio (S/N) was high, the deepest H and \ion{He}{1} lines are present, 
and the spectrum is free of Earth's atmospheric lines.  Radial velocities were measured from 
the cross-correlation maxima with uncertainties estimated from the maximum
likelihood method of \citet{Zucker2003}.  The mean spectrum of Regulus and the model
spectrum used as the cross-correlation template are shown in Figure~1. 

\begin{figure}[ht] 
\begin{center} 
{\includegraphics[angle=90,height=11cm]{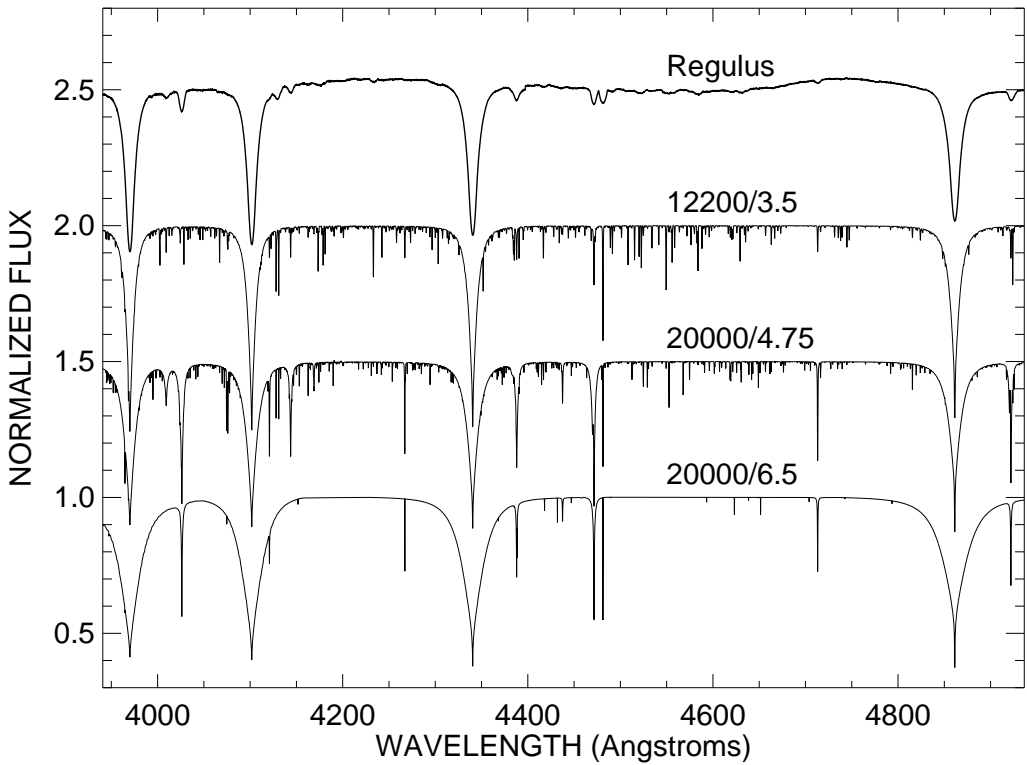}} 
\end{center} 
\caption{The mean, normalized spectrum of Regulus (offset by +1.5) together with model spectra used 
for calculating the cross-correlation functions.  The BLUERED model for the primary (offset by +1.0) 
is plotted second from top and is based upon assumed parameters $T_{\rm eff}=$ 12.2~kK,
$\log g = 3.5$, $V\sin i = 0$, and solar abundances. The BSTAR2006 model for the secondary (offset by +0.5) 
appears third from top and is based upon assumed parameters $T_{\rm eff}=$ 20~kK,
$\log g = 4.75$, $V\sin i = 0$, and solar abundances. The TMAP model for the secondary
is plotted at the bottom for $T_{\rm eff}=$ 20~kK, $\log g = 6.5$, $V\sin i = 0$, and solar abundances.
} 
\label{fig1} 
\end{figure}

\placefigure{fig1}     
 
\placetable{tab1}      
\begin{deluxetable}{cccccl} 
\tablewidth{0pc} 
\tabletypesize{\scriptsize} 
\tablenum{1} 
\tablecaption{Journal of Spectroscopy \label{tab1}} 
\tablehead{ 
\colhead{Run} & 
\colhead{Dates} & 
\colhead{Range} & 
\colhead{Resolving Power} & 
\colhead{} & 
\colhead{Observatory/Telescope/} \\ 
\colhead{Number} & 
\colhead{(BY)} & 
\colhead{(\AA)} & 
\colhead{($\lambda/\triangle\lambda$)} & 
\colhead{$N$} & 
\colhead{Spectrograph}} 
\startdata 
16\dotfill & 2009.9 -- 2013.9 & 4400 -- 7500               &\phn   50000 &   17 & Okayama/1.8 m/HIDES \\ 
17\dotfill & 2010.3 -- 2013.9 & 3620 -- 8530               &\phn   50000 &\phn3 & Cerro Amazones/1.5 m/BESO \\ 
18\dotfill & 2006.2 -- 2008.0 & 3643 -- 7990               &\phn   55000 &\phn4 & Keck I/10 m/HIRES \\ 
19\dotfill & 2014.3           & 1160 -- 3046               &      114000 &\phn3 & HST/2.4 m/STIS \\ 
20\dotfill & 2005.4 -- 2016.3 & 3700 -- 10000              &\phn   65000 &   25 & CFHT/3.6 m/ESPaDOnS and TBL/2.0 m/NARVAL \\
\enddata 
\end{deluxetable} 
 
These 52 new radial velocity measurements are collected in Table~2 that lists 
the heliocentric Julian date of mid-observation, the orbital phase (see below), 
the radial velocity and its uncertainty, the observed minus calculated residual (see below), 
and the run number corresponding the observational journal in Table~1.  
The velocities given in Table~2 for the ESPaDOnS/NARVAL set (run \#20) are
offset by $-3.7$ km~s$^{-1}$ in order to match the systemic velocity found 
from other measurements (see below).  We found that some of the uncertainties
based on the standard deviation of a small set of nightly averages were 
unrealistically small, so we set a lower limit of 1 km~s$^{-1}$ in such cases.
On the other hand, the uncertainties for the ESPaDOnS/NARVAL set were slightly 
too large and led to a reduced $\chi^2<1$ in the fit, so these were rescaled 
to obtain a final reduced $\chi^2$ equal to one. 
 
\placetable{tab2}      
\begin{deluxetable}{lccccc} 
\tabletypesize{\scriptsize} 
\tablewidth{0pt} 
\tablenum{2} 
\tablecaption{Radial Velocity Measurements \label{tab2}} 
\tablehead{ 
\colhead{Date}              & 
\colhead{Orbital}           & 
\colhead{$V_r$}             & 
\colhead{$\sigma (V_r)$}    & 
\colhead{$(O-C)$}           & 
\colhead{Run}               \\ 
\colhead{(HJD$-$2,400,000)} & 
\colhead{Phase}             & 
\colhead{(km s$^{-1}$)}     & 
\colhead{(km s$^{-1}$)}     & 
\colhead{(km s$^{-1}$)}     & 
\colhead{Number}            } 
\startdata 
 53509.741 \dotfill & 0.967 &      \phs $11.1$ &  0.8 &      $-0.7$ & 20 \\
 53511.763 \dotfill & 0.018 &      \phs $11.7$ &  0.8 &      $-0.2$ & 20 \\
 53806.881 \dotfill & 0.377 & \phn \phs $ 0.1$ &  1.2 & \phs $ 1.2$ & 18 \\
 53903.755 \dotfill & 0.792 & \phn \phs $ 6.5$ &  0.8 & \phs $ 0.1$ & 20 \\
 53904.750 \dotfill & 0.817 & \phn \phs $ 8.1$ &  0.8 & \phs $ 0.6$ & 20 \\
 53905.748 \dotfill & 0.842 & \phn \phs $ 8.7$ &  0.8 & \phs $ 0.2$ & 20 \\
 54086.035 \dotfill & 0.338 & \phn \phs $ 1.7$ &  2.7 & \phs $ 1.3$ & 18 \\
 54110.576 \dotfill & 0.950 &      \phs $12.3$ &  0.9 & \phs $ 0.7$ & 20 \\
 54250.868 \dotfill & 0.448 & \phn \phs $ 0.2$ &  1.0 & \phs $ 3.0$ & 18 \\
 54461.132 \dotfill & 0.691 & \phn \phs $ 3.9$ &  3.2 & \phs $ 2.2$ & 18 \\
 55191.627 \dotfill & 0.907 &      \phs $10.3$ &  1.0 &      $-0.4$ & 16 \\
 55193.627 \dotfill & 0.957 &      \phs $10.3$ &  1.0 &      $-1.3$ & 16 \\
 55199.518 \dotfill & 0.104 & \phn \phs $ 8.8$ &  1.0 &      $-1.7$ & 16 \\
 55231.164 \dotfill & 0.893 & \phn \phs $ 8.1$ &  1.0 &      $-2.2$ & 16 \\
 55275.009 \dotfill & 0.986 &      \phs $11.2$ &  1.0 &      $-0.7$ & 16 \\
 55304.614 \dotfill & 0.725 & \phn \phs $ 7.0$ &  1.0 & \phs $ 3.8$ & 17 \\
 55338.040 \dotfill & 0.558 & \phn      $-4.4$ &  1.0 &      $-1.7$ & 16 \\
 55343.486 \dotfill & 0.694 & \phn \phs $ 0.5$ &  1.9 &      $-1.3$ & 17 \\
 55519.381 \dotfill & 0.080 &      \phs $11.9$ &  1.0 & \phs $ 0.9$ & 16 \\
 55520.359 \dotfill & 0.105 &      \phs $11.0$ &  1.0 & \phs $ 0.6$ & 16 \\
 55521.370 \dotfill & 0.130 &      \phs $10.4$ &  1.0 & \phs $ 0.8$ & 16 \\
 55597.085 \dotfill & 0.018 &      \phs $11.7$ &  1.0 &      $-0.2$ & 16 \\
 55640.406 \dotfill & 0.098 & \phn \phs $ 9.9$ &  0.9 &      $-0.7$ & 20 \\
 55656.375 \dotfill & 0.496 & \phn      $-3.7$ &  1.0 &      $-0.5$ & 20 \\
 55658.431 \dotfill & 0.548 & \phn      $-3.3$ &  0.9 &      $-0.5$ & 20 \\
 55694.955 \dotfill & 0.458 & \phn      $-4.2$ &  1.0 &      $-1.3$ & 16 \\
 55751.777 \dotfill & 0.875 &      \phs $10.5$ &  0.8 & \phs $ 0.8$ & 20 \\
 55755.759 \dotfill & 0.975 &      \phs $12.7$ &  0.8 & \phs $ 0.8$ & 20 \\
 55757.755 \dotfill & 0.024 &      \phs $12.7$ &  0.8 & \phs $ 0.8$ & 20 \\
 55842.366 \dotfill & 0.134 &      \phs $10.4$ &  1.0 & \phs $ 1.0$ & 16 \\
 55906.287 \dotfill & 0.728 & \phn \phs $ 4.8$ &  1.0 & \phs $ 1.5$ & 16 \\
 55906.673 \dotfill & 0.738 & \phn \phs $ 3.0$ &  1.0 &      $-0.8$ & 20 \\
 55932.158 \dotfill & 0.373 & \phn      $-0.5$ &  0.8 & \phs $ 0.4$ & 20 \\
 55935.119 \dotfill & 0.447 & \phn      $-2.2$ &  0.8 & \phs $ 0.5$ & 20 \\
 55935.652 \dotfill & 0.460 & \phn      $-4.4$ &  1.0 &      $-1.5$ & 20 \\
 55936.583 \dotfill & 0.484 & \phn      $-4.3$ &  1.0 &      $-1.2$ & 20 \\
 55960.115 \dotfill & 0.071 &      \phs $11.8$ &  0.8 & \phs $ 0.6$ & 20 \\
 56055.069 \dotfill & 0.438 & \phn      $-2.8$ &  1.0 &      $-0.1$ & 16 \\
 56104.787 \dotfill & 0.678 & \phn \phs $ 1.9$ &  0.7 & \phs $ 0.8$ & 20 \\
 56105.794 \dotfill & 0.703 & \phn \phs $ 2.9$ &  0.8 & \phs $ 0.7$ & 20 \\
 56106.798 \dotfill & 0.728 & \phn \phs $ 4.3$ &  0.8 & \phs $ 1.0$ & 20 \\
 56107.797 \dotfill & 0.753 & \phn \phs $ 5.1$ &  0.8 & \phs $ 0.6$ & 20 \\
 56274.249 \dotfill & 0.904 &      \phs $13.2$ &  1.0 & \phs $ 2.6$ & 16 \\
 56615.296 \dotfill & 0.408 & \phn \phs $ 2.1$ &  1.0 & \phs $ 4.0$ & 16 \\
 56623.875 \dotfill & 0.622 & \phn      $-3.1$ &  1.5 &      $-2.0$ & 17 \\
 56625.250 \dotfill & 0.657 & \phn \phs $ 5.1$ &  1.0 & \phs $ 4.9$ & 16 \\
 56752.788 \dotfill & 0.837 & \phn \phs $ 7.2$ &  4.0 &      $-1.1$ & 19 \\
 56754.680 \dotfill & 0.884 & \phn \phs $ 7.7$ &  5.6 &      $-2.4$ & 19 \\
 56756.703 \dotfill & 0.935 & \phn \phs $ 7.3$ &  2.2 &      $-4.1$ & 19 \\
 57460.468 \dotfill & 0.484 & \phn      $-5.2$ &  0.9 &      $-2.0$ & 20 \\
 57468.597 \dotfill & 0.687 & \phn \phs $ 0.1$ &  0.8 &      $-1.4$ & 20 \\
 57485.509 \dotfill & 0.108 & \phn \phs $ 8.4$ &  0.9 &      $-1.8$ & 20 \\
\enddata 
\end{deluxetable} 
 
We solved for the orbital elements using the {\tt rvfit} 
code\footnote{http://www.cefca.es/people/$^\sim$riglesias/rvfit.html} written by
\citet{Iglesias-Marzoa2015}. Initial fits yielded zero eccentricity within 
errors, so we adopted a circular orbital solution.  The ESPaDOnS/NARVAL set
of measurements is particularly useful because they are numerous, cover 
a time span of 11 y, and the spectra were made with essentially identical instruments. 
Thus, we began with a fit of the ESPaDOnS/NARVAL set alone that is given in column 3 of   
Table~3, which lists the orbital period $P$, epoch of maximum radial velocity $T_0$,
semiamplitude $K_1$, and systemic velocity $V_0$, along with the derived 
projected semimajor axis $a_1\sin i$, mass function $f(M)$, and root-mean-square
(r.m.s.) of the residuals from the fit.  These elements agree within uncertainties 
with those derived earlier (Table 3, column 2; \citealt{Gies2008}) with the exception of the systemic 
velocity.  This difference may result from the properties of the ESPaDOnS/NARVAL
wavelength calibration, the adopted spectral normalization of the echelle 
blaze function in the vicinity of the strong Balmer lines, and/or asymmetries in the 
cross-correlation functions related to mismatch of the model and observed spectra.  
In any case, we decided to apply a constant offset of $-3.7$ km~s$^{-1}$ to the ESPaDOnS/NARVAL
velocities that was based upon a comparison with $V_0$ from a preliminary solution 
that used all the other velocities.  The final column of Table~3 gives the elements
from the complete merged set of radial velocities from \citet{Gies2008} and Table~2, 
and the associated radial velocity curve is shown in Figure~2.  The revised elements
from all the data agree with the earlier results, and the enlarged data set achieves
lower uncertainties for the orbital elements. 

\begin{figure}[h] 
\begin{center} 
{\includegraphics[angle=0,height=10cm]{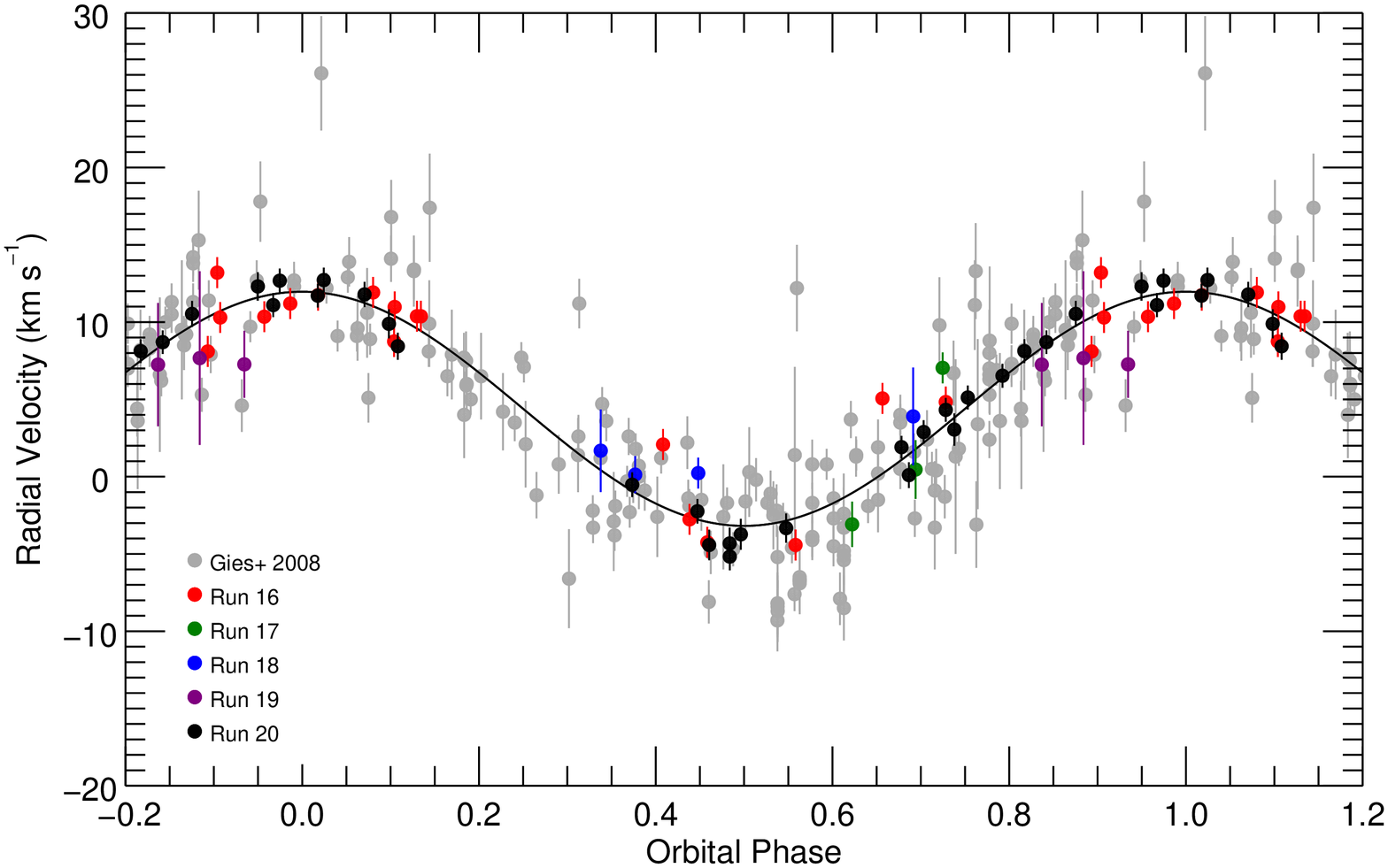}} 
\end{center} 
\caption{The observed and fitted radial velocity curves based upon 
all the available spectroscopy.  Symbol colors are associated with 
the observational sets listed in Table 1.} 
\label{fig2} 
\end{figure}

\placefigure{fig2}     
 
\placetable{tab3}      
 \begin{deluxetable}{lccc} 
\tablewidth{0pc} 
\tablenum{3} 
\tablecaption{Circular Orbital Elements\label{tab3}} 
\tablehead{ 
\colhead{Element}                 &\colhead{Gies et al. (2008)}&\colhead{ESPaDOnS/NARVAL}&\colhead{All data}} 
\startdata 
$P$ (d)                              \dotfill & $40.11 \pm 0.02$        & $40.101 \pm 0.008$  & $40.102 \pm 0.002$  \\ 
$T_0$ (HJD -- 2,400,000) \dotfill & $44526.3 \pm 0.3$     & $57480.7 \pm 0.4$     & $57481.2 \pm 0.2$  \\ 
$K_1$ (km s$^{-1}$)         \dotfill & $7.7 \pm 0.3$             & $7.9 \pm 0.2$             & $7.58 \pm 0.12$  \\ 
$V_0$ (km s$^{-1}$)         \dotfill & $4.3 \pm 0.2$             & $7.9 \pm 0.2)$            & $4.39 \pm 0.09$ \\ 
$a_1\sin i$ ($R_\odot$)    \dotfill & $6.1 \pm 0.3$             & $6.3 \pm 0.2$             & $6.00 \pm 0.17$ \\ 
$f(M)$ ($M_\odot$)           \dotfill & $0.0019 \pm 0.0002$ & $0.0021 \pm 0.0002$ & $0.00181 \pm 0.00015$ \\ 
r.m.s. (km s$^{-1}$)           \dotfill & 2.8                              & 0.8                              & 3.3 \\
\enddata 
\end{deluxetable} 

\newpage

 
\section{Detection of the Pre-White Dwarf Spectrum}     

\subsection{Detection Method}
The companion of Regulus is very faint, so it is extremely 
difficult to identify its spectral lines directly even in very high S/N spectra. 
The best approach is to find ways to average the spectral signal of the companion
using many spectral features and including all the spectra of interest. 
Creating a cross-correlation function (CCF) of an observed and model 
spectrum effectively multiplexes all the matching spectral features into 
a single ``superline'' profile.  Thus, our search method involves 
an analysis of the CCFs that are shifted into the frame of reference
of the companion star and then co-added to obtain the best S/N ratio. 

The ESPaDOnS/NARVAL collection of spectra is ideal for 
this purpose because the individual and nightly averaged spectra have very 
high S/N ($\approx 1000$ in the better exposed parts) 
and they are well distributed over the range of orbital Doppler shifts (Fig.~2). 
We began by calculating the CCFs with a model spectrum for the
assumed parameters of the companion.  The flux-normalized model was
taken from the BSTAR2006 grid from \citet{Lanz2007} and transformed 
to the same $\log \lambda$ wavelength scale used for all the 
ESPaDOnS/NARVAL spectra.  The model spectrum is based upon an assumed 
solar metallicity, a given secondary effective temperature and gravity, 
and no rotational broadening.  We caution that the pre-white dwarf companion 
may have a photospheric composition that differs from the solar one 
and that the highest gravity available in the BSTAR2006 grid $\log g =4.75$
is probably much lower than the actual value (Section 3.6).  Older white dwarf stars 
have very diverse abundances in their outer atmospheres, but the 
companion of Regulus is probably too young (Section 4) for elemental 
diffusion processes to have altered the abundances significantly. 
Thus, a model spectrum based upon solar abundances 
is a reasonable starting assumption in a search for the spectral 
line patterns of the companion in the spectrum of Regulus. 

The CCFs derived using a model spectrum with $T_{\rm eff}=20$~kK (Section 3.4)
are dominated by the correlation of the spectral features with the 
very broad lines in the much brighter spectrum of Regulus itself 
(compare the observed and model spectra in Fig.~1).  Consequently, the next 
step is to make a first-order correction to remove the broad part of the CCF 
function that is due to the primary star.  All 25 CCFs were shifted in 
velocity space using the orbital solution for the ESPaDOnS/NARVAL spectra
(Table 3, column 3) to place them into the primary's frame of reference, 
and then they were co-added to form a primary CCF component.  This step
assumes that any secondary contribution is too weak and too spread over the 
velocity range of the secondary between observations to have any impact on the 
shape of the primary CCF.  Once calculated, we shifted the average primary CCF  
to the observer's frame for each observed orbital phase and then subtracted 
it from the observed CCF to derive a matrix of CCF residuals.  
The CCF residuals are dominated by spectrum to spectrum variations due to 
small differences in flux rectification that produce low frequency 
variations with a full sample standard deviation of $\sigma = 0.008$
(approximately $1\%$ of the average CCF maximum before subtraction). 
 
It is very difficult to discern the secondary signal in any 
individual residual CCF.  The residuals after subtraction of the primary 
component are dominated by low frequency, small amplitude 
variations that are probably related to small differences between spectra 
in the final flux placement after echelle blaze normalization.  
Thus, before coadding the secondary-shifted, residual CCFs, we subtracted 
from each one a Gaussian smoothed version of itself (with a Gaussian FWHM = 
460 km~s$^{-1}$, selected to be much larger than the expected widths of most of the 
secondary's spectral features).  This effectively removed the low frequency 
variations while maintaining the sharper, high frequency parts of the CCFs.
We experimented with different amounts of Gaussian filtering for background subtraction,
and in general a larger Gaussian FWHM results in incomplete removal of the
background variations (making it harder to discern the companion signal) 
while a smaller Gaussian FWHM tends to remove the signal from the companion. 

The final steps are to shift each residual CCF into the velocity frame 
of the companion (for an assumed companion semiamplitude $K_2$), 
coadd the residuals from each observation into a mean CCF, 
and then subtract a linear fit of the background of the mean CCF 
far from the CCF central velocity in order to compare any resulting 
signal with a baseline set to zero.  
We show the result of this procedure in Figure~3 that shows the mean 
of the CCF residuals in the velocity frame of the companion.  
There is a central peak in the mean CCF that represents the 
sought-after signal from the spectrum of the faint companion star.  
There are significant variations away from the central peak that are 
probably related to correlation with echelle order boundaries 
and other systematic imperfections in the spectra.  Instead of 
using these variations to estimate the CCF noise level, we performed 
a numerical experiment by making 1000 iterations of a bootstrap
random selection of 25 phase samples from the set of CCFs, 
and then calculating the mean CCF for each sample in the same way as 
described above. The horizontal dotted lines show the $\pm 1 \sigma$ 
standard deviation levels in the final mean derived from the 
bootstrap procedure.  The vertical dotted line marks the position of 
the central peak that occurs close to the expected zero velocity location
and reaches about $5\sigma$ above the noise level. 

\placefigure{fig}     
\begin{figure}[h] 
\begin{center} 
{\includegraphics[angle=90,height=10cm]{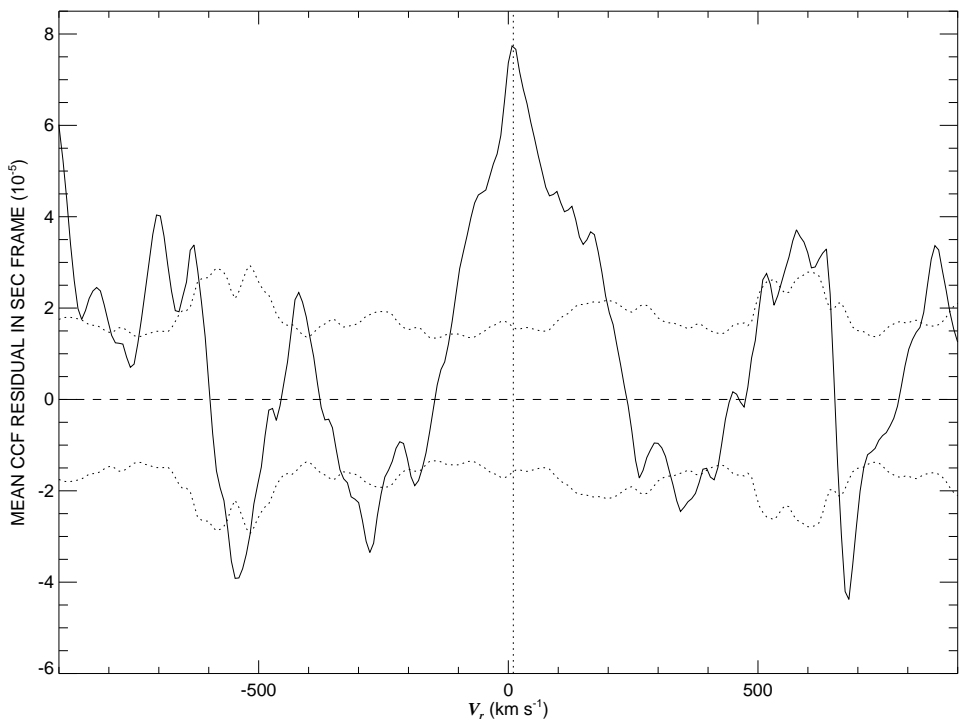}} 
\end{center} 
\caption{The mean of the CCF residuals after removal of the component for 
the spectrum of the Regulus primary and after shifting into the velocity 
reference frame of the companion star 
(for an adopted companion semiampitude of $K_2 = 91$ km~s$^{-1}$).  
The horizontal dashed and dotted lines indicate the locations of 
the mean linear background and the $\pm 1\sigma$ levels from 
a bootstrap analysis, respectively.  
The vertical dotted line marks the measured peak location of a central signal 
formed by correlation with the companion's spectral features.} 
\label{fig3} 
\end{figure} 
 
 \subsection{Validation Tests}
 
 The signal detected corresponds to a very faint companion star spectrum 
 (Section 3.4), so it is important to consider tests of the significance of the 
 CCF peak feature.  We present several numerical tests here that were 
 suggested to us by the referee and that help to validate the reality of 
 the detected signal.  The first is the construction of a ``skew map'' 
 representation of the orbital phase-integrated signal that has proved to
 be a valuable tool to detect the faint spectral features of cool mass-donor
 stars in cataclysmic variables 
 \citep{Smith1993,Smith1998,VandePutte2003}.
 The concept of the skew map is to evaluate the line integral in a set
 of CCFs in a matrix of velocity shift and orbital phase for a source that
 exhibits Doppler shifts according to
 \begin{equation}
 V_r (\phi) = V_0 - V_x \cos (2 \pi \phi) + V_y \sin (2 \pi \phi)
 \end{equation}
 where $\phi$ is the sinusoidal representation of orbital phase 
 (equal to our adopted cosinusoidal phase in Table 3 plus 0.75).
 $V_x$ represents axial motion directed from primary towards secondary
 and $V_y$ represents orthogonal orbital motion for a circular orbit. 
 An actual signal from an orbiting companion only occurs for $V_x=0$ and 
 $V_y>0$, but it is helpful to calculate the skew map for the full range 
 in $V_x$ and $V_y$ in order to judge the strength of a companion 
 signal with other spurious peaks for non-physical parameters. 
 
 We constructed such a skew map by shifting each of the residual CCFs 
 according to the test values of $(V_x, V_y)$ and the observed orbital 
 phases for each of the 25  ESPaDOnS/NARVAL spectra.  
 Then the line integral was taken as the average of the shifted CCF 
 residuals at the velocity of the peak of the signal shown in Figure~3. 
 We found that the background variations of the CCFs varied in a 
 systematic way with parameter $V_y$.  Thus instead of subtracting 
 a smoothed CCF background as done in Section 3.1, we formed a 
 CCF background from the CCF residuals at offset positions of 
 $V_y \pm 70$ km~s$^{-1}$ for each value of $V_y$, and then we 
 subtracted this background from the CCF before evaluating the 
 mean value at  the peak position.  This procedure may also remove
 the sought after signal if the companion spectral features have 
 half-widths similar to or larger than 70 km~s$^{-1}$, but the 
 correlation widths for the spectrum of the companion are much
 smaller than this offset (Section 3.3).

\placefigure{fig}     
\begin{figure}[h!] 
\begin{center} 
{\includegraphics[angle=0,height=10cm]{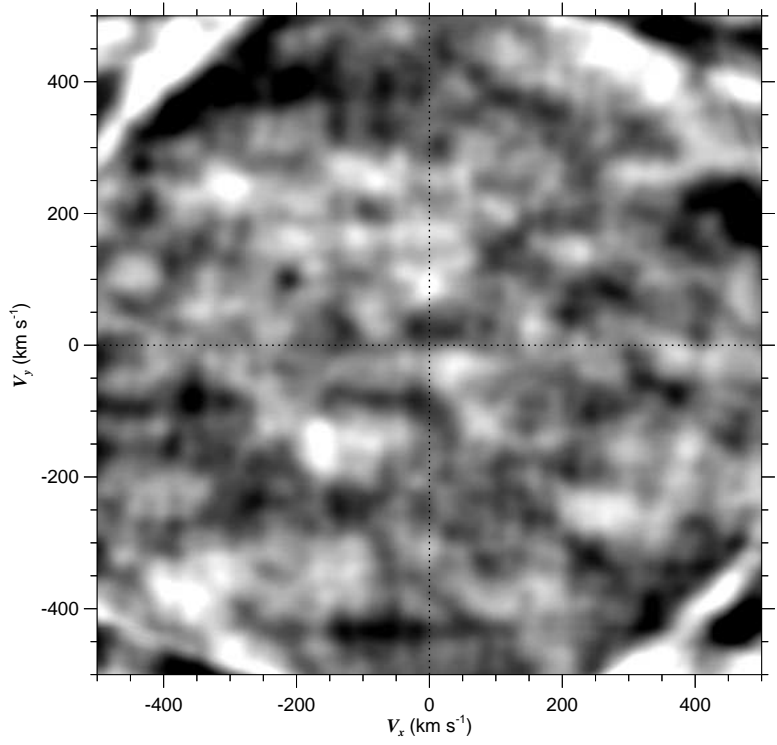}} 
\end{center} 
\caption{A skew map of the CCF residuals plotted as functions of orbital velocity
vectors $(V_x, V_y)$.  The gray intensity indicates the average peak  CCF residual 
strength after $(V_x, V_y)$ velocity shifting at each orbital phase and 
subtraction of background trends estimated from offset $V_y$ positions.
The black to white intensity corresponds to $\pm1.6 \times 10^{-5}$ in CCF value.
The only valid companion signal is located along $V_x=0$ and $V_y > 0$, and 
the detected signal is the bright spot located at $(V_x, V_y) = (0,91)$ km~s$^{-1}$.} 
\label{fig4} 
\end{figure} 
 
 The background subtracted skew map from the CCF residuals is 
shown as a grayscale image in Figure~4.  This image shows that 
indeed there are other spurious peaks as well as arc features 
at the extremes of the diagram that arise from edge effects in the 
CCFs.   However, in the region of interest of $(V_x, V_y) = (0, >0)$
there is only one bright feature that corresponds to the signal 
revealed in Figure~3 (at $(V_x, V_y) = (0, 91)$ km~s$^{-1}$).
Thus, the predominance of this signal at a location in the skew map 
where it is expected to occur offers support for a true detection. 
 
 The second validation test is based upon the idea that the detected 
 peak results from assuming coherent Doppler shifts for the companion star. 
 If instead the orbital phases and velocities were incorrect, then any 
 peak in the co-added residuals would be very weak because the 
 false Doppler shifts would spread the signal over a range in velocity. 
 Therefore, we conducted a numerical test by randomly scrambling the 
 assigned orbital phases of the observations and then forming the 
 residual CCFs over a range of assumed $K_2 = 10$ to 200 km~s$^{-1}$.
 We then determined the resulting CCF maximum near the velocity of 
 the peak shown in Figure~3 over the grid of trial $K_2$ values.  
 We made 10000 iterations of the procedure
 and collected the resulting peak maxima for statistical analysis. 
 We show in Figure~5 the probability distribution of finding a peak 
 with a height $h$ or larger from the randomly scrambled set of CCFs.
 The vertical line in Figure~5 gives the observed height of the peak
 we identify using the correctly ordered phases. The chance of 
 finding a peak this high is very small, about  1 in 3000 for the random 
 sampling of the observed phases (making this a $4\sigma$ detection). 
 
\placefigure{fig}     
\begin{figure}[h] 
\begin{center} 
{\includegraphics[angle=0,height=10cm]{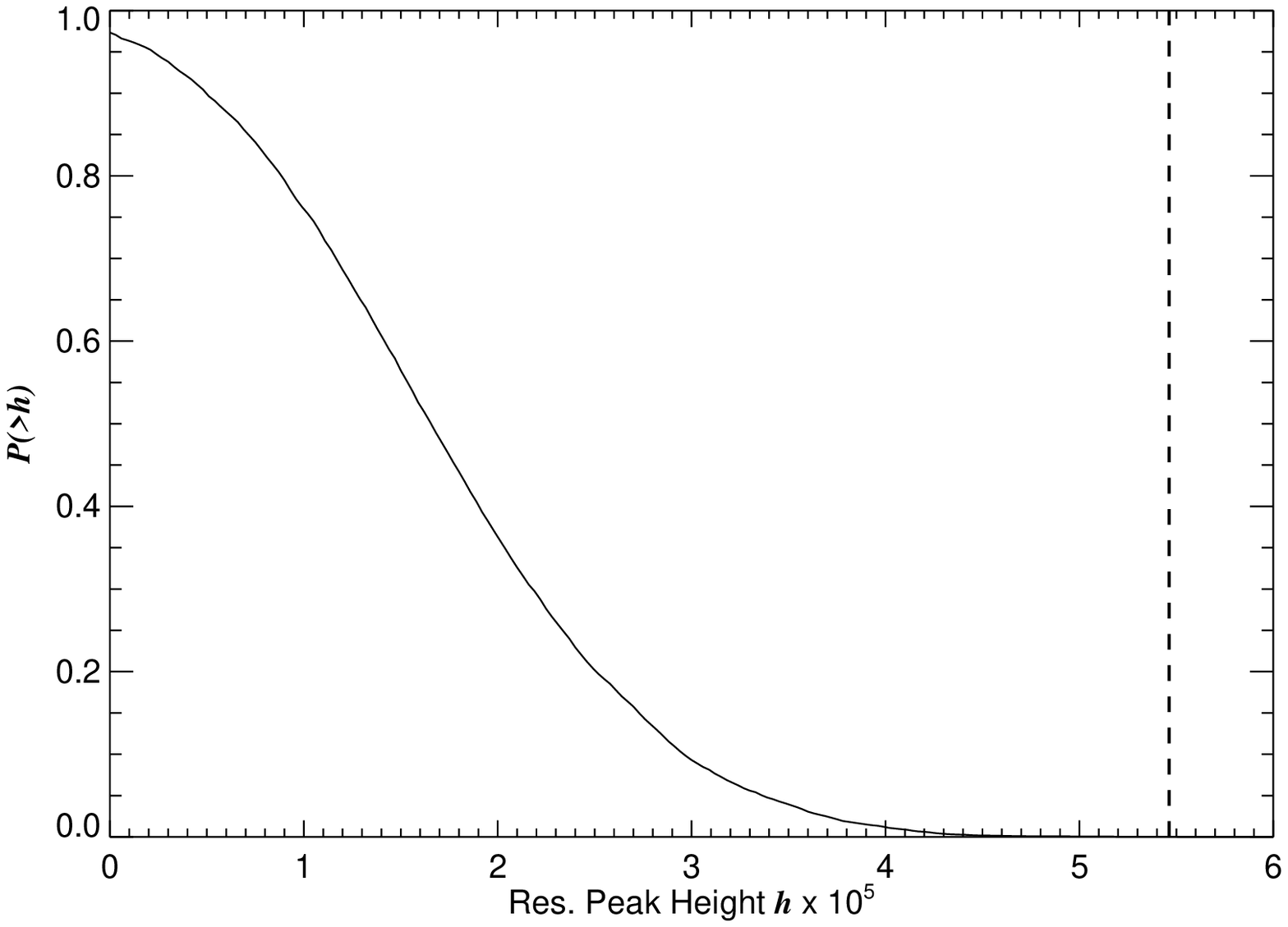}} 
\end{center} 
\caption{The probability of finding a maximum CCF residual with peak height
greater than $h$ based upon numerical experiments made by randomly 
scrambling the assigned orbital phases.  The vertical dashed line indicates 
the observed peak height of the detected signal for the companion. 
} 
\label{fig5} 
\end{figure} 
 
The third validation test involved inserting an artificial spectrum into the
observed spectra with pre-defined Doppler shifts and flux ratio, and 
then using the same CCF procedures to detect the artificial signal. 
This test demonstrates how the performance of the method deteriorates 
for ever fainter companion spectra.  The simulation used the same 
TLUSTY model companion spectrum  for the inserted signal, 
but the assumed semiamplitude was assumed to be much larger 
($K_2 = 300$ to 500 km~s$^{-1}$) in order to avoid overlapping the 
actual and simulated CCF signals.   The model spectra $S_m$ were added 
to the observed spectra $S_o$ to form combined spectra $S_c$, 
\begin{equation}
S_c = (S_o + r S_m) / (1 + r)
\end{equation}
where the assumed monochromatic flux ratio is $r=f_2/f_1$.   Then the 
combined model spectra were used to form CCF residuals as done for the 
observations, and a search was made for the corresponding peak in 
the co-added residuals over a grid of trial $K_2$ values to find
the semiamplitude that maximizes the signal (Section 3.3). 

Removal of the background variations in the CCFs was again 
accomplished by smoothing each CCF and subtracting the 
smoothed version to reveal the high frequency component 
of the correlation.  We found, however, that the final signal 
was still dwarfed by background variations for faint flux ratios,
so for this test each co-added CCF for a given test $K_2$ 
was compared to that for the $r=0$ case of no companion 
spectrum insertion to determine the net peak strength. 

This simulation was done for 16 flux ratios on a logarithmic grid, 
and for each flux ratio ten simulations were made by inserting 
the model spectra into the observed set for the Doppler shifts 
associated with a randomly selected value of $K_2$ between 
300 and 500 km~s$^{-1}$.  The pairs of input and derived $K_2$ 
were collected from the simulations, and finally the standard 
deviation of their difference was determined for each bin of 
model flux ratio. 

The results of the simulation are shown in Figure~6 that shows 
the standard deviation between input and derived semiamplitude 
as a function of model flux ratio.  The agreement is good and 
the standard deviation is low for brighter model companions.
On the other hand, as expected the derived semiamplitude departs 
from the model value at progressively lower flux ratios.
The vertical dashed line shows the estimated monochromatic flux ratio 
for the companion of Regulus (Section 3.4), and this occurs at
a value that is about a factor of 10 larger than that where the 
deviations between model and derived semiamplitude become 
relatively large.  Thus, this simulation indicates that the companion 
is bright enough that the spectra S/N and CCF analysis are sufficient 
for a reliable detection and for an estimate of semiamplitude (Section 3.3).
 
\placefigure{fig}     
\begin{figure}[h] 
\begin{center} 
{\includegraphics[angle=0,height=10cm]{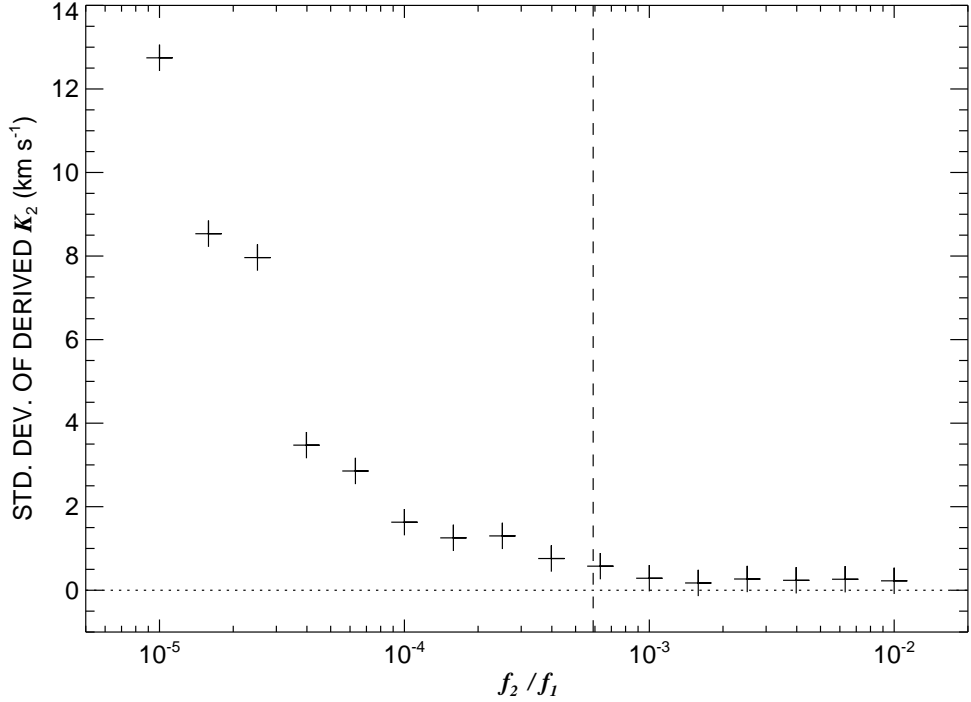}} 
\end{center} 
\caption{The standard deviation between given and derived estimates of 
the companion semiamplitude $K_2$ as a function of assumed flux ratio
$f_2/f_1$ from numerical simulations with insertion of a model secondary 
spectrum into the observed set.  The vertical dashed line indicates the 
derived flux ratio for the companion (Section 3.4).
} 
\label{fig6} 
\end{figure} 
 
All three of these validation tests indicate that the CCF signal we 
detect can be confidently associated with the spectral correlation from 
the companion of Regulus.  In the next subsections, we assume 
that CCF signal does reveal the flux of the pre-white dwarf star, 
and we use the observed CCF properties to estimate physical 
parameters for this star. 

\pagebreak
 
 \subsection{Secondary Orbital Semiamplitude}
 
Tests show that the strength and sharpness of the central peak varies with the 
assumed semiamplitude $K_2$, becoming increasingly weaker, broader, and unfocussed 
as the trial value of $K_2$ departs from its actual value.  This means that we can 
perform trial runs over a grid of $K_2$ test values to find a best-fit estimate 
that maximizes the peak strength and minimizes the net width of the residual peak.  
We found that many of the peaks and troughs surrounding the central peak 
grew in amplitude with increasing assumed $K_2$, indicating that they are 
artifacts introduced by correlation with stationary imperfections in the spectra. 
Consequently, instead of using the simple subtraction of a smoothed version
of the CCF, we adopted the mean formed from an unrealistically large 
$K_2 = 200$ km~s$^{-1}$ shift-and-add of the CCFs as a reference for the background. 
For each trial value of $K_2$ from 10 to 200 km~s$^{-1}$, this background template 
was scaled by a factor of $K_2/(200 {\rm ~km~s}^{-1})$ and then subtracted from the 
observed residual CCF to isolate the peak from the companion alone. 
The local background was set interactively around the central peak 
for each trial value of $K_2$, and then the maximum peak height above 
the local background was recorded together with the second moment of the
peak distribution as a measurement of the peak half width.  The measured 
peak height and half width are plotted as functions of $K_2$ in Figure~7.

\begin{figure}[h] 
\begin{center} 
{\includegraphics[angle=0,height=10cm]{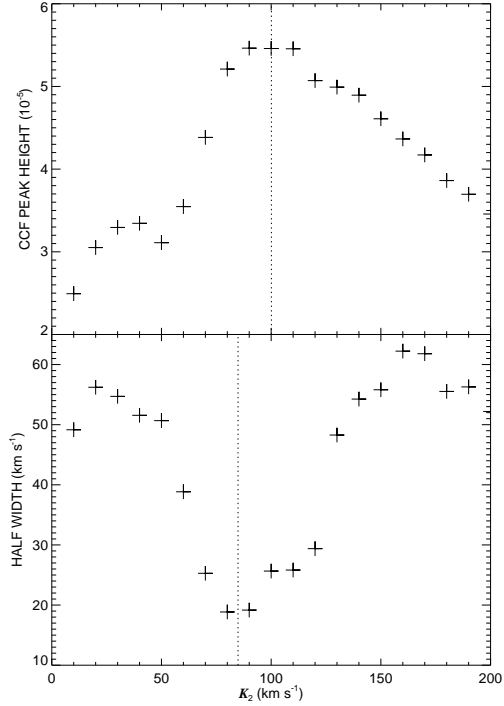}} 
\end{center} 
\caption{The derived peak height (top) and half width (bottom) of 
the central peak in the subtracted mean CCF for shift-and-add 
velocities associated with a grid of assumed companion semiamplitude $K_2$.
The dotted lines indicate the best fit for the companion semiamplitude
derived from the peak height ($K_2 = 100 \pm 19$ km~s$^{-1}$) and the half-width
($K_2 = 85 \pm 15$ km~s$^{-1}$) in the top and bottom panels,
respectively.} 
\label{fig7} 
\end{figure}

\placefigure{fig7}     
 
We next implemented a bootstrap selection of 25 non-unique spectra 
from the original set in order to repeat the analysis using the same 
local background velocity regions that were selected interactively. 
The same procedure of forming the co-added CCF residuals was run 
over a grid of test $K_2$ values to find the value that maximized the 
peak height and that minimized the peak half-width.  We performed 
3000 bootstrap iterations of observation resampling to form 
distribution functions for the best fitting $K_2$ values, and we determined 
uncertainties from the standard deviations of these distributions.
The peak height criterion attains a maximum at $K_2 = 100 \pm 19$ km~s$^{-1}$,
and the peak half-width is a minimum (narrowest) at $K_2 = 85 \pm 15$ km~s$^{-1}$.
We take the error weighted means of the semiamplitude and uncertainty 
to derive our final estimate, $K_2 = 91 \pm 16$ km~s$^{-1}$.

This result implies a mass ratio $M_2/M_1 = K_1 / K_2 = 0.083 \pm 0.015$ 
that is consistent with expectations from the mass function and 
evolutionary scenarios \citep{Gies2008, Rappaport2009}
and with similar results for stripped down companions in 
Be binaries \citep{Wang2018} and 
EL~CVn binaries \citep{Maxted2014, Matson2015, Wang2020}.
Using the orbital solution from all spectra (Table~3, column 4) leads to 
mass product estimates of $M_1 \sin ^3 i = (3.7 \pm 1.4) M_\odot$ and 
$M_2 \sin ^3 i = (0.31 \pm 0.10) M_\odot$.  
Regulus presents a spin inclination axis of $i=86.3$ deg 
to our line of sight \citep{Che2011}, 
and this should be equal to the orbital inclination according to models
of donor spin-up through binary mass transfer.  Thus, we expect that 
$\sin ^3 i\approx 1$, so the actual masses are close to the 
$M\sin^3 i$ products given above. 

The minimum peak half-width of 20 km~s$^{-1}$ is close to the value  
expected from cross-correlation with a model spectrum with no rotation,
so any rotational broadening in the companion spectrum is small and unresolved.
Models of interacting binaries indicate that the mass donor will grow to 
the dimensions of its Roche lobe, and the subsequent tidal torques will force
the entire star into synchronous rotation.  This would imply that the 
companion rotation period is the same as the orbital period (40 d), so 
any projected rotational velocity would be insignificant and the spectral 
lines would appear narrow.  Synchronous rotation and sharp lines are found
for the stripped companions in many other pre-white dwarf plus main sequence 
star binaries \citep{Matson2015,Wang2018,Wang2020}.  

\subsection{Dependence of Peak Height on Temperature and Flux Ratio}

The CCF peak height is also related to the degree of spectral feature matching 
between the companion and model spectra.  We repeated the CCF calculations 
using a range in model effective temperature and constructing models from 
the UVBLUE \citep{Rodriguez2005} and BSTAR2006 \citet{Lanz2007} grids for
$T_{\rm eff}$ less than and greater than 15~kK, respectively.  
The peak strength reaches a maximum for $T_{\rm eff} = (20 \pm 4)$~kK, and the 
peak disappears for models with $T_{\rm eff} < 8$~kK and $T_{\rm eff} > 40$~kK.
Thus, the companion appears to be much hotter than the primary star, 
which has $T_{\rm eff} = 12.2$~kK \citep{Gies2008}.  

The absolute residual peak height is directly related to the 
companion flux contribution to the combined spectrum.  Thus, we can 
infer the monochromatic flux ratio $f_2/f_1$ from the observed signal strength
by determining the predicted relationship between these parameters. 
We calculated this functional relation by performing a series of simulations 
of composite spectra over a grid of assumed flux ratios. 
For each assumed flux ratio, we created a simulated set of 25 spectra 
by adding the TLUSTY model spectrum of the companion (Fig.~1) to the observed 
spectrum for the assumed companion Doppler shifts and flux contribution. 
We formed cross-correlation functions with this simulated set, and
then followed the same procedure to remove the primary's contribution and 
co-add the the residuals in the secondary's frame as we did with the 
actual observations.  The final step was to measure the net growth of 
the signal peak height as a function of assumed $f_2/f_1$ and form
the linear relation between these parameters.   The observed peak 
appears to be superimposed upon a local maximum in Figure~5, so we 
used the net peak strength above the local background of  
$(3.2\pm 1.0) \times 10^{-5}$ together with the derived linear relation to estimate 
the flux ratio, $f_2/f_1 = (5.9 \pm 1.8)\times 10^{-4}$.  This is equivalent 
to a $B$-band magnitude difference of $\triangle m = 8.1 \pm 0.4$ mag.

This observed monochromatic flux ratio $f_2/f_1$ is related to the ratio of stellar radii by
$${f_2 / f_1} = (R_2 / R_1)^2 (F_2/F_1)$$
where $R$ is the radius and $F$ is the emitted monochromatic flux.  
We calculated that the emitted flux ratio per unit area is 
$F_2/F_1 = 2.19$ for the $V$-band and $T_1=12.2$~kK and $T_2=20$~kK
using model fluxes from \citet{Bertone2008}. 
Thus, the derived radius ratio from the relation above is $R_2 / R_1 = 0.016 \pm 0.003$. 
The error estimate is based on both the uncertainty in $f_2/f_1$ from the 
peak height and the uncertainty in $F_2/F_1$ from the range in possible 
companion temperature and associated model flux.
\citet{Che2011} measured the polar and equatorial radii of Regulus, 
and using their average of $R_1/R_\odot = 3.7$, 
we arrive at a companion radius of $R_2 / R_\odot = 0.061\pm 0.011$.  

\subsection{Gravitational Redshift}

Compact stars like white dwarfs are dense enough that 
their spectrum suffers a significant gravitational 
redshift in many cases \citep{Greenstein1967, Parsons2017}.  The prediction in 
this case is a velocity redshift of $0.636 ~(M/M_\odot)/(R/R_\odot)$ km~s$^{-1}$
that is equal to $+3.2 \pm 1.2$ km~s$^{-1}$ for the companion mass and radius 
given above.  A parabolic fit of the central peak velocity in Figure~3 
yields a similar value of $+10 \pm 5$ km~s$^{-1}$, where the uncertainty 
is dominated by the assumed value for the systemic velocity (taken from 
the ESPaDOnS/NARVAL solution in Table~3).  Thus, the slight offset of 
the peak in Figure~5 is consistent with our expectation of the gravitational 
redshift.  

\subsection{Spectral Template Properties}

The results above indicate that the detected CCF peak has the 
properties expected for a stripped down companion star.  
On the other hand, the spectrum template used to derive the CCFs 
is based on a gravity of $\log g = 4.75$ (the largest in the 
TLUSTY model grid) while our results imply a larger gravity of 
$\log g = 6.4$.  There are only a few published model synthetic 
spectra of solar abundance atmospheres for gravities this large.
We show one example in Figure~1 for a model spectrum with 
$T_{\rm eff} = 20$~kK and $\log g =6.5$ from the 
T\"{u}bingen NLTE Model Atmosphere Package (TMAP)
\citep{Werner2003,Rauch2003,Rauch2018} 
that we obtained from the Spanish Virtual Observatory 
web site\footnote{http://svo2.cab.inta-csic.es/theory/newov2/}.
This model is based upon a NLTE treatment of the energy levels
for H and He, and also includes line opacity for C, N, O, Ne, 
Na, and Mg.  It is similar in appearance to the TLUSTY model 
for $T_{\rm eff} = 20$~kK and $\log g =4.75$ (Fig.~1) except 
that the H Balmer line wings have greater pressure broadening. 
Note that the Balmer line profiles in both the TLUSTY and TMAP
models display very narrow absorption cores, and we suspect 
that the narrow appearance of the final CCF peak (Fig.~3) 
is due to the correlation of the Balmer line cores in the 
observed and model spectra.  Note that our method of background 
subtraction will remove correlation from broad features, so 
the subtracted CCF lacks the broader component from correlation 
with the Balmer line wings.  We found that the CCF peak of 
the companion is also present when the TMAP spectrum is used
as a template, but the peak is somewhat weaker.  

We also explored calculating the CCFs using the short and long 
wavelength halves of the spectra separately (divided at a 
mid-point wavelength of 4372 \AA ).  We found that the 
peak was strong for the short wavelength part but nearly absent 
for the long wavelength part. The short wavelength region
contains the strong Balmer lines of H$\gamma$, H$\delta$, and
H$\epsilon$, while the long wavelength section only contains
H$\beta$ and weaker lines of \ion{He}{1} $\lambda\lambda 
4387, 4471, 4713, 4921$.  The difference in the CCF strength 
for the two regions suggests that the H Balmer lines are 
the dominant contributor to the CCF amplitude.  
In the similar, lower mass binary, EL~CVn, evidence suggests the 
presence of a thick hydrogen envelope even after extreme 
mass loss \citep{Maxted2014}, and the spectrum of 
the companion in EL~CVn shows strong H Balmer lines with narrow 
cores \citep{Wang2020}, consistent with their appearance in 
the model CCF template spectra for solar abundances that 
we use here.  Thus, the companion of Regulus probably has a 
photosphere with a significant hydrogen abundance. 
 
\section{Discussion}                                

The cross-correlation analysis of the large set of high S/N 
spectra from ESPaDOnS/NARVAL reveals a weak signal that has 
all the properties expected for the spectral signature of the 
pre-white dwarf companion of Regulus. The CCF peak only appears using
model spectra for a hot object, and it attains the strongest and narrowest 
morphology for a secondary semiamplitude that is consistent with the 
predicted extreme mass ratio and low mass of the pre-white dwarf remnant. 
The narrow appearance of the peak agrees with expectations of a low 
projected rotational velocity for synchronous rotation that is predicted
in binary evolution models.  The mass and radius implied by 
the observations lead to a predicted small gravitational redshift that
is consistent with the measured peak velocity.  Taken together, these  
results indicate that CCF peak marks the first detection of the 
spectral features of the pre-white dwarf companion of Regulus. 

The stellar masses we derive from the CCF analysis are 
similar to those proposed in earlier work \citep{Gies2008}.  
\citet{Che2011} used the stellar temperature and luminosity and the Yale-Yonsei 
evolutionary tracks to find a primary mass of $M_1 = (4.15 \pm 0.06) M_\odot$. 
This agrees within uncertainties with our result, $M_1 = (3.7 \pm 1.4) M_\odot$.  

The pre-white dwarf star has a low mass, $M_2 = (0.31 \pm 0.10) M_\odot$,
that represents the mass of the surviving core after completion of mass transfer. 
There is a close relationship between core mass and the properties 
of the former, Roche-filling red giant, and this implies that 
the core mass is related to the orbital period of the system. 
\citet{Rappaport2009} describe how the relation applies to the 
Regulus system and how the orbital period leads to a predicted 
pre-white dwarf mass of $M_2 = (0.28 \pm 0.05) M_\odot$, and this 
estimate agrees within uncertainties with the measured mass.  

The evolution of the stellar core after envelope removal was 
investigated by \citet{Istrate2016} using MESA models to study the 
internal structure and tracks in the $(\log T_{\rm eff}, \log L/L_\odot)$ 
plane.  We show in Figure~8 one example for a He core pre-white dwarf with 
a mass of $0.305 M_\odot$ from the basic set they present, i.e., 
without consideration of elemental diffusion and rotation.  
The remnant first increases in temperature at constant luminosity, and 
then begins to fade towards the white dwarf cooling sequence.  
The star may experience episodic and short hydrogen shell flashes 
that cause excursions to higher luminosity.  We also plot the properties
of the pre-white dwarf in the Regulus binary in Figure~8, and it appears
near the hot end of the cooling sequence predicted by the model.  
The diamond and plus sign tick marks indicate time intervals of 1 and 10 Myr, 
respectively, in the evolutionary tracks, and this model predicts that the 
remnant will have a temperature and luminosity like those observed between 
3 and 100 Myr from the time of envelope removal. 

\begin{figure}[h!] 
\begin{center} 
{\includegraphics[angle=90,height=10cm]{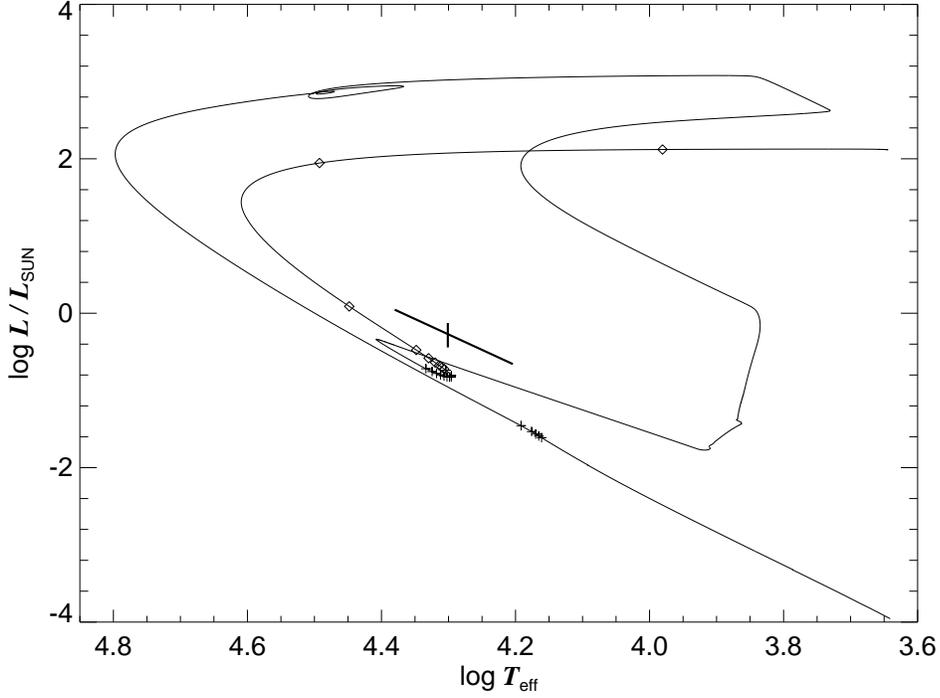}} 
\end{center} 
\caption{The evolutionary track in the $(\log T_{\rm eff}, \log L/L_\odot)$ 
plane of a $0.305 M_\odot$ stripped core from the work of \citet{Istrate2016}.
Diamonds mark the first nine Myr since removal of the envelope 
(in increments of 1 Myr), and plus signs mark subsequent ages from 
10 to 150 Myr (in increments of 10 Myr).  The inclined error bars show 
the estimated parameters and uncertainties for the pre-white dwarf 
companion of Regulus.} 
\label{fig8} 
\end{figure}

\placefigure{fig8}     
 
Regulus itself was probably spun up to its fast rotation rate around 
the same time as the peak of the mass transfer rate that led to the 
stripping of the envelope of the mass donor.  According to the 
companion model track in Figure~8, this occurred some 3 to 100 Myr ago. 
The upper range is similar to the age derived for the primary from \citet{Che2011}
of 90 Myr based upon evolutionary tracks.  Their estimate presumably 
refers to the time since the mass gainer was rejuvenated as a larger
mass star by mass transfer from the mass donor, and it agrees with 
the elapsed time since the conclusion of mass transfer of $\approx 90$ 
Myr in the evolutionary model by \citet{Rappaport2009}. These estimates are a 
significant fraction of the main sequence lifetime of a $4.15 M_\odot$
star (about 160 Myr; see Fig.~7 in \citealt{Che2011}).  
Thus, stars spun up by mass transfer may appear 
as rapid rotators for a significant part of their H-core burning lifetime.  
This suggests that many of the rapid rotators among the early-type stars may 
have also experienced spin-up through binary mass transfer \citep{deMink2014}. 
If so, then there are many other cases awaiting discovery of faint white dwarfs 
and their progenitors orbiting rapidly rotating stars. 

\pagebreak

Our new results suggest a different interpretation of the occultation 
of Regulus by the asteroid Adorea \citep{Dunham2017}.  
According to the orbital elements presented in Table 3, probable masses, 
and the distance from {\it Hipparcos}, the angular semimajor axis of the
orbit is $\approx 15$ milliarcsec (mas).  The projected separation at the time 
of the occultation was $\approx 8$ mas, much smaller than the 
angular diameter of the asteroid ($\approx 58$ mas) that 
transited nearly centered over Regulus from the observing site. 
This indicates that both Regulus and its companion were occulted by Adorea. 
Furthermore, the mid-occultation magnitude of $V=12.4 \pm 0.5$ is 
significantly fainter than the estimated companion magnitude derived 
from the monochromatic flux ratio, 
$V_2=V_1({\rm Regulus})-2.5 \log (f_2/f_1)=1.4+8.1=9.5$ mag.  
Consequently, it appears that 
flux recorded during the occultation was not that of the companion,
but was probably the faint halo of diffracted light of Regulus itself. 
Based upon the observing site location and circumstances of the occultation, 
we estimate that the angular separation from the Regulus to the limb 
of Adorea varied from 0 to 18 mas with a time average of 11 mas.  
Using the program {\tt Occult} (v4.6.0) written by 
Dave Herald\footnote{http://www.lunar-occultations.com/iota/occult4.htm}, 
this corresponds to a decrease of 11.2 mag for the asymptotic part of the 
Fresnel diffraction pattern. Thus, the diffracted flux would yield a 
a Regulus magnitude of $V = 1.4 + 11.2 = 12.6$ on average during the event.
If we add the flux of Adorea ($V=14.0$), then the total magnitude would be
$V=12.3$, the same as observed within uncertainties.  
This suggests that the observed magnitude during the occultation 
is consistent with the estimated diffraction flux of Regulus
plus the flux of Adorea, and hence, there is no need to posit 
flux from another source.  Nevertheless, future occultation observations
may yet lead to a detection of the companion in favorable circumstances. 
For example, \citet{Richichi2011} used VLT observations of lunar 
occultations to detect binaries as close as 7 mas in separation with 
a magnitude difference as large as 6.5 mag. 

The small angular separation of the stars means that direct imaging 
will remain challenging.  However, given the equator-on orientation
of the spin axis of Regulus, it is likely that the orbital axis 
has a similiar inclination.  For the stellar radii and semimajor 
axis found above, eclipses should occur for an orbital inclination 
$i>87.6$ deg, which is slightly larger than the spin inclination found 
by \citet{Che2011}, $i=86.3^{+1.0}_{-1.6}$ deg.  Thus, it is 
marginally possible that the system exhibits shallow eclipses.  
The deeper (shallower) eclipse would occur at orbital phase 0.75 (0.25) in 
our ephemeris, and create dips of about 0.6 (0.3) millimag (for total eclipses) 
with a duration less than 17 hours.
\citet{Rucinski2011} attempted a search for eclipses with the 
{\it MOST} satellite, but unfortunately, the noise in the light 
curve was comparable to the expected eclipse depth during the 
time of the secondary minimum when they observed (phase 0.25). 
Future space-based photometry might yet detect the eclipses of 
this remarkable system. 

 
\acknowledgments 
 
This material is based upon work supported by the 
National Science Foundation under Grant No.~AST-1908026. 
We thank Eiji Kambe and Juliette Becker for sharing their 
radial velocity data on Regulus in advance of publication. 
We also thank an anonymous referee for very helpful comments. 
The Okayama Astrophysical Observatory data were obtained during the 
commissioning phase observations of the fiber-fed system.
Based in part on observations made with NASA/ESA Hubble Space Telescope, 
obtained from the Mikulsky Archive at Space Telescope Science Institute, 
operated by the Association of Universities for Research in Astronomy, Inc., 
under NASA contract NAS 5-26555.  Support for ASTRAL is provided by grants
HST-GO-12278.01-A and HST-GO-13346.01-A from STScI. 
This work is based in part on observations obtained
with ESPaDOnS at the Canada-France-Hawaii Telescope
(CFHT) and with NARVAL at the T\'{e}lescope Bernard Lyot (TBL). 
CFHT/ESPaDOnS is operated by the National Research Council of Canada, 
the Institut National des Sciences de l'Univers of the Centre National 
de la Recherche Scientifique (INSU/CNRS) of France, and the University of
Hawaii, while TBL/NARVAL is operated by INSU/CNRS.
GAW acknowledges support from the Natural Sciences and Engineering 
Research Council of Canada (NSERC) in the form of a Discovery Grant.
Institutional support has been provided from the GSU College 
of Arts and Sciences.  We are grateful for all this support. 
 
 
\facilities{HST (STIS), CFHT (ESPaDOnS), Keck:I (HIRES), OAO:1.88m (HIDES), OCA:Hexapod (BESO), TBL (NARVAL)}

\software{rvfit \citep{Iglesias-Marzoa2015}, fxcor \citep{Alpaslan2009}}


\bibliographystyle{aasjournal}
\bibliography{paper}{}
 
\end{document}